\documentclass[sigconf]{acmart}
\AtBeginDocument{%
  }

\usepackage{amsmath}
\usepackage{tikz}
\usepackage{subcaption}
\usepackage{pifont}
\usepackage{siunitx}
\usepackage{lipsum}
\usepackage{colortbl}
\usepackage{enumitem}
\usepackage{xcolor}
\usepackage{makecell}
\usepackage{algorithm}
\usepackage{algpseudocode}
\usepackage{graphicx}
\usepackage{booktabs}
\usepackage{multirow}
\usepackage{soul}
\usepackage{chngcntr}
\usepackage{array,adjustbox,tabularx}
\usepackage{subcaption}

\newcolumntype{Y}{>{\centering\arraybackslash}X}
\definecolor{myblue}{RGB}{79,173,234}

\AtBeginDocument{%
  }
\setcopyright{acmlicensed}
\copyrightyear{2026}
\acmYear{2026}
\acmDOI{XXXXXXX.XXXXXXX}

\newcommand{\squishlist}
{\begin{itemize}[itemsep=1pt,parsep=2pt,topsep=3pt,partopsep=0pt,leftmargin=0em, itemindent=1em,labelwidth=1em,labelsep=0.5em]}
\newcommand{\squishend}{\end{itemize}}
\newcommand{\squishenum}{\begin{enumerate}[itemsep=1pt,parsep=2pt,topsep=3pt,partopsep=0pt,leftmargin=0em, itemindent=1.5em,labelwidth=1em,labelsep=0.5em]}
\newcommand{\squishsubenum}{\begin{enumerate}[itemsep=1pt,parsep=2pt,topsep=0pt,partopsep=0pt,leftmargin=0em,listparindent=1.5em,labelwidth=1em,labelsep=0.5em]}
\newcommand{\squishenumend}{\end{enumerate}}

\newcommand{\red}[1]{\textcolor{black}{#1}}

\newcommand{\sysname}{\textit{WeeCare}}
\makeatletter

\copyrightyear{2026}
\acmYear{2026}
\setcopyright{cc}
\setcctype{by}
\acmConference[ISWC '26]{Proceedings of the 2026 ACM International Symposium on Wearable Computers}{October 11--15, 2026}{Shanghai, China}
\acmBooktitle{Proceedings of the 2026 ACM International Symposium on Wearable Computers (ISWC '26), October 11--15, 2026, Shanghai, China}
\acmDOI{10.1145/3830727.3834809}
\acmISBN{979-8-4007-2872-3/2026/10}

\begin{document}



\newcommand{\weecare}{\includegraphics[height=1em]{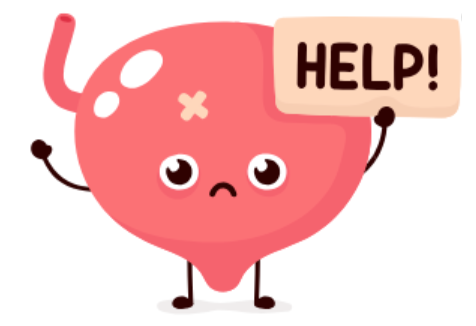}}

\title{{\sysname}: Towards Handheld Bladder Fullness Sensing\\with a Conformable Pad\\}


\author{Zhikai Qin}
\affiliation{
  \institution{Carnegie Mellon University\country{}}
}

\author{Siqi Zhang}
\affiliation{
  \institution{Carnegie Mellon University\country{}}
}

\author{Shuyi Zeng}
\affiliation{
  \institution{Carnegie Mellon University\country{}}
}

\author{Xiyuxing Zhang}
\affiliation{
  \institution{Carnegie Mellon University\country{}}
}

\author{Junyi Zhu}
\affiliation{
  \institution{University of Michigan\country{}}
}

\author{Justin Chan}
\affiliation{
  \institution{Carnegie Mellon University\country{}}
}

\renewcommand{\shortauthors}{Qin et al.}

\begin{teaserfigure}
\centering
\includegraphics[width=.9\linewidth]{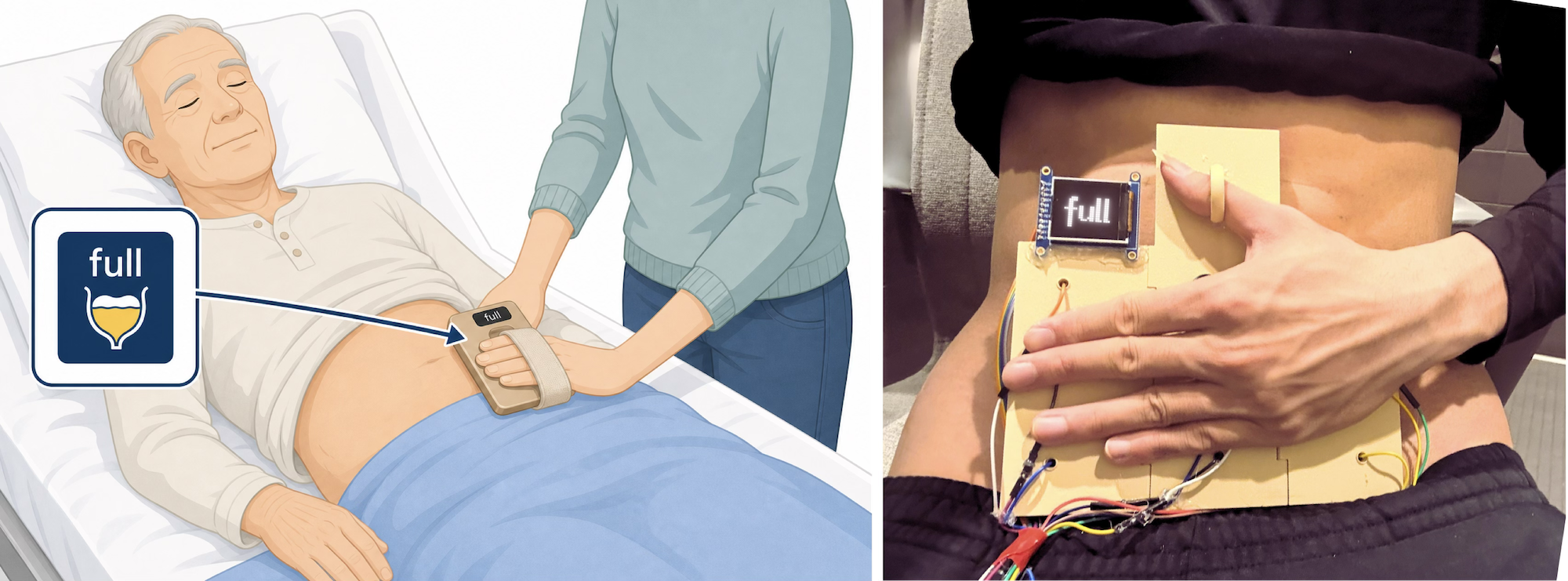}
\vspace{-1em}
\caption{{\sysname}, a \red{proof-of-concept} handheld electrical impedance tomography pad for on-demand bladder fullness sensing \red{in a controlled setting}. (Left) Envisioned use in a care setting for patients with bladder dysfunction who lose the sensation of bladder filling. On-demand bladder fullness sensing has the potential to reduce the need for frequent and unnecessary catheterization. (Right) {\sysname} prototype, a foldable grid of fabric electrodes that conforms to the abdomen, with an integrated display.}
\Description{Description of figure}
\label{fig:concept_ai}
\end{teaserfigure}

\begin{abstract}
Patients with bladder dysfunction often lose the sensation of bladder fullness and cannot void naturally, forcing reliance on fixed-schedule catheterization that is uncomfortable and risks complications. We present {\sysname}, a handheld conformable pad with fabric electrodes for on-demand bladder fullness sensing using electrical impedance tomography (EIT). The central challenge is that repeated removal and reattachment can introduce variation in electrode position and contact quality. We assess {\sysname} along three axes: in-silico simulations characterizing electrode layout and noise robustness, in-vitro phantom experiments across urine salinities and filling levels, 
\red{and an in-vivo study tracking voiding dynamics and fullness sensing across 8 participants, with filling dynamics characterized in a single participant. Our results provide an early assessment of {\sysname}'s feasibility under controlled conditions.}

\end{abstract}

\maketitle

\section{Introduction}
Patients with bladder dysfunction often cannot reliably sense bladder fullness or voluntarily void~\cite{dorsher2012neurogenic,samson2007neurogenic,verpoorten2008neurogenic}. This affects individuals with spinal cord injury or spina bifida, where damaged neural pathways impair urinary function~\cite{dorsher2012neurogenic,samson2007neurogenic,verpoorten2008neurogenic}. Clean intermittent catheterization is the gold standard for emptying the bladder~\cite{verpoorten2008neurogenic,dorsher2012neurogenic,lamin2016clean}. However, because patients may have a decreased or absent urge to urinate, catheterization is performed on a fixed schedule, every 2 to 4 hours, rather than when the bladder is actually full~\cite{dorsher2012neurogenic}.

The problem is that catheterizing too early and frequently is uncomfortable, increases caregiver burden, and raises the risks of complications such as urinary tract infections, urethral bleeding, and bladder stones~\cite{igawa2008catheterization,wyndaele2002complications}. Conversely, if the bladder overfills before the next scheduled catheterization, it can distend beyond its safe capacity, potentially causing urinary tract damage~\cite{madersbacher1990various,verpoorten2008neurogenic}.

Conventionally, ultrasound bladder scanners~\cite{bvi9400,primeplus,ultrasoundClinic2005} are used to non-invasively estimate bladder volume but they are typically expensive and bulky with a rigid and uncomfortable probe. Wearable, miniaturized monitors using ultrasound~\cite{wearablrUntrsound2026,WearableUntrasound2024,lee2024intelligent}, near-infrared spectroscopy ~\cite{molavi2013noninvasive,fechner2023near}, or bioelectrical impedance sensing~\cite{nonConsistent2011,artefact2023,schlebusch2014bladder}, which adhere to the abdomen~\cite{toymus2024integrated} or integrate into belts~\cite{noyori2022non} and undergarment~\cite{zhang2023privee}, have been proposed for continuous monitoring, but wearing such systems all day can be uncomfortable.

This gap motivates our research question: \textit{can bladder fullness be sensed non-invasively, comfortably, and on demand?} If successful, such an approach could allow patients to check bladder fullness before deciding whether to catheterize, 
\red{and could potentially fit into care workflows where a caregiver takes on-demand measurements across multiple patients.}

Here, we present {\sysname}, a handheld conformable pad for on-demand bladder fullness sensing using electrical impedance tomography. Since urine is conductive, changes in bladder volume alter the conductivity of the pelvic region. Electrical impedance tomography (EIT) captures these changes by injecting small electrical currents through surface electrodes and measuring the resulting voltages to reconstruct an internal conductivity map of the bladder~\cite{schlebusch2014bladder}.

However, the design of a handheld EIT device also introduces a significant challenge which is that repeated removal and application of the device to the skin introduces noise and positional variability~\cite{eitkit2021} that interfere with bladder measurements. We investigate whether bladder fullness can still be obtained through repeated reattachment using a handheld pad along three dimensions.

\textit{First,} we develop an in-silico simulation to evaluate the robustness of EIT bladder fullness and volume estimation under changes in contact quality and position across different electrode layouts. \textit{Second,} we develop a benchtop phantom model of the bladder and perform in-vitro measurements under repeated removal and reattachment, evaluating volume estimation across urine salinities and bladder filling levels. 
\red{\textit{Finally}, as an initial proof-of-concept, we conduct in-vivo measurements on 8 participants. We show the system tracks voiding dynamics and fullness sensing across participants, and, in a single participant, demonstrate filling dynamics. We frame this as a feasibility demonstration in a controlled lab setting rather than clinical validation.}

\section{Background and related work}

\begin{figure}[t]
\includegraphics[width=\linewidth]{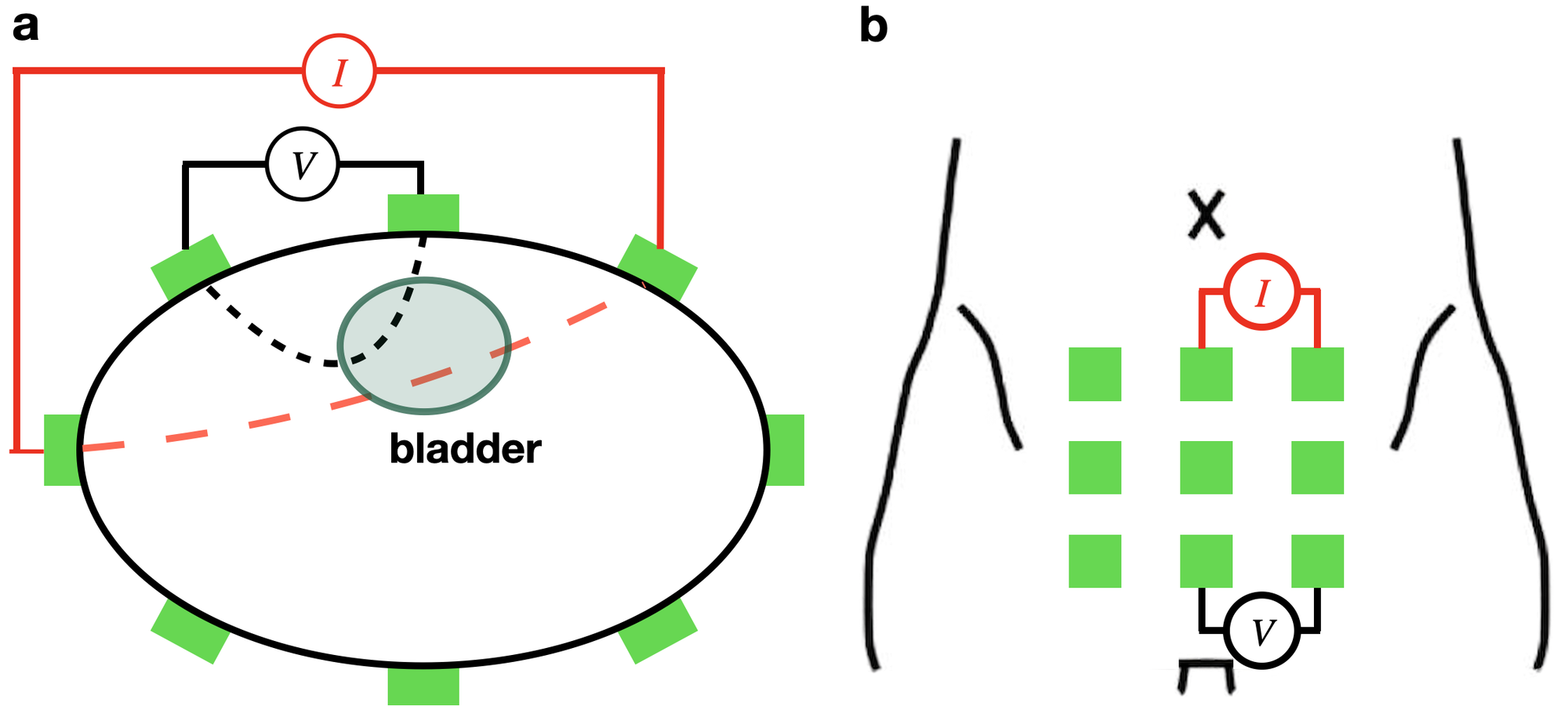}
\vspace{-2em}
\caption{(a) Conventional ring electrode layout in belt form factor. (b) {\sysname} planar grid electrode layout in handheld pad form factor.}
\label{fig:comparison}
\end{figure}

\noindent {\bf Bioimpedance and electrical impedance tomography.}
Bioelectrical impedance sensing measures changes in tissue conductivity when the tissue is exposed to a bio-electrical current with a single global impedance measurement. Given the conductivity of urine, it has been used to estimate bladder filling and volume~\cite{stablewearble2025,artefact2023,BIAfd2017,zhang2023privee}. 

Electrical impedance tomography (EIT) extends this by reconstructing spatial conductivity variations from multi-electrode measurements~\cite{tfdEIT2025,carbonImage2018,eitBook2005,review2022}. As shown in Fig.~\ref{fig:comparison}a, prior EIT work places electrodes in a ring that encircles the torso, typically embedded in a belt. Current is injected and voltage measured between electrode pairs around this ring, so the sensing field wraps fully around the abdomen and maximizes spatial coverage~\cite{schlebusch2014bladder,tfdEIT2025}. 


Because the belt stays fixed on the body for an entire filling cycle, these designs can report quantitative bladder volume~\cite{schlebusch2014bladder,artefact2023}. However, such a design can be sensitive to patient weight and can be uncomfortable to wear for long periods~\cite{artefact2023,absoluteEIT2022}.

In contrast, {\sysname} uses a planar grid of electrodes in a handheld pad (Fig.~\ref{fig:comparison}b). All electrodes sit on a single patch of skin over the bladder rather than encircling the body, so the device can be applied and removed on demand. However, the flexibility of repeated removal and replacement introduces variability in electrode positioning and contact noise between measurements. For our in-vivo testing, we target fullness sensing, a binary empty-versus-full decision, 
\red{which can inform catheterization timing.}

\noindent {\bf Near-infrared spectroscopy.} Optical approaches such as near-infrared spectroscopy (NIRS)~\cite{NIR2008,molavi2013noninvasive,fong2018restoring}, have been used to estimate urine quantity based on the absorption properties of near-infrared light in human tissue and water using LEDs and photoedetectors at the abdomen but are sensitive to tissue variability and have shown limited accuracy in practice.

\noindent {\bf Ultra-wideband.} Ultra-wideband radar has been proposed for contactless measurement of bladder volume~\cite{li2010performance,o2013bladder,krewer2014development}. The bladder is illuminated with a UWB pulse and differences in the dielectric properties of urine and the surrounding tissue are detected in the reflections as the bladder fills. However, these works have so far been validated only in simulation and phantom studies, and performance is sensitive to device position and orientation.
\section{{\sysname}}

Our system design involves three phases: 
\squishenum
\item {\bf In-silico.} We perform a simulation-based design exploration to assess how electrode layout, noise, and non-ideal positioning affect system performance, which informs hardware design.
\item {\bf In-vitro.} We conduct benchtop phantom validation to assess the effects of repeated pad removal and replacement, as well as urine salinity, which can vary with hydration and fluid intake.
\item {\bf In-vivo.} We evaluate our bladder fullness sensing system on human participants to characterize its feasibility under realistic physiological and repeated-placement conditions.
\squishenumend

\subsection{Simulation-based design exploration}

\subsubsection{Simulation setup}\leavevmode

\noindent {\bf Torso and bladder model.} To guide the design of our hardware system, we performed a simulation of our proposed planar electrode grid layout in PyEIT~\cite{liu2018pyeit} to study how electrode layout affect the reconstructed EIT image. \red{We model the abdomen with a custom 18,362-element torso mesh~\cite{kiwibunn_torso_mesh} and the bladder as an ellipsoid.} 

\begin{figure}[t]
\includegraphics[width=\linewidth]{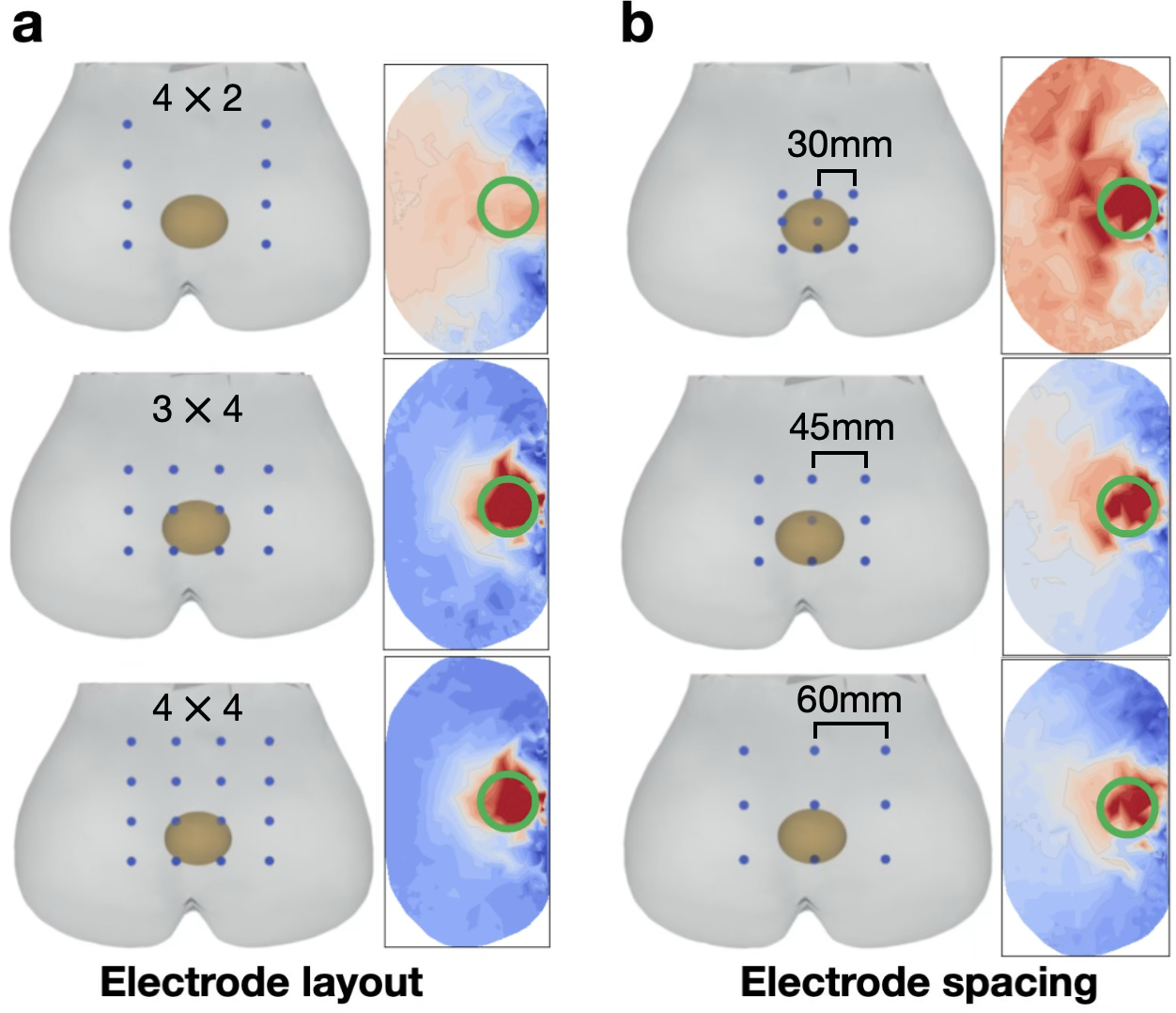}
\vspace{-2em}
\caption{Simulation illustrating the effect of (a) electrode layout and (b) electrode spacing on EIT image reconstruction for a bladder filled to 100~mL. The figures show a 3D rendering of the abdomen and filled bladder, the electrode positions, and the 2D conductivity image reconstruction, {slicing at the middle of the bladder}. The green circle indicates the location of the bladder.}
\label{fig:sim_electrodes}
\vspace{-1em}
\end{figure}



\noindent {\bf Electrode channel selection.} An EIT system produces \red{many voltage readings}, \red{but because a bladder conductivity change alters the entire abdominal field, these channels are highly correlated and redundant,} so we down-select electrodes to a smaller subset. In a $3\times3$ grid of $N=9$ electrodes with 4-pole sensing, each channel uses four electrodes: one pair for current injection and one for voltage measurement (Fig.~\ref{fig:comparison}). \red{From the full set of 756 four-pole channels, we down-select to 48: 36 axis-aligned rectangular channels (formed by injecting current across one edge of a row/column rectangle and measuring across the opposite edge) and 12 diagonal channels (with injection and sensing pairs placed along the grid diagonals).}

\begin{figure}[t]
    \centering
\includegraphics[width=\linewidth]{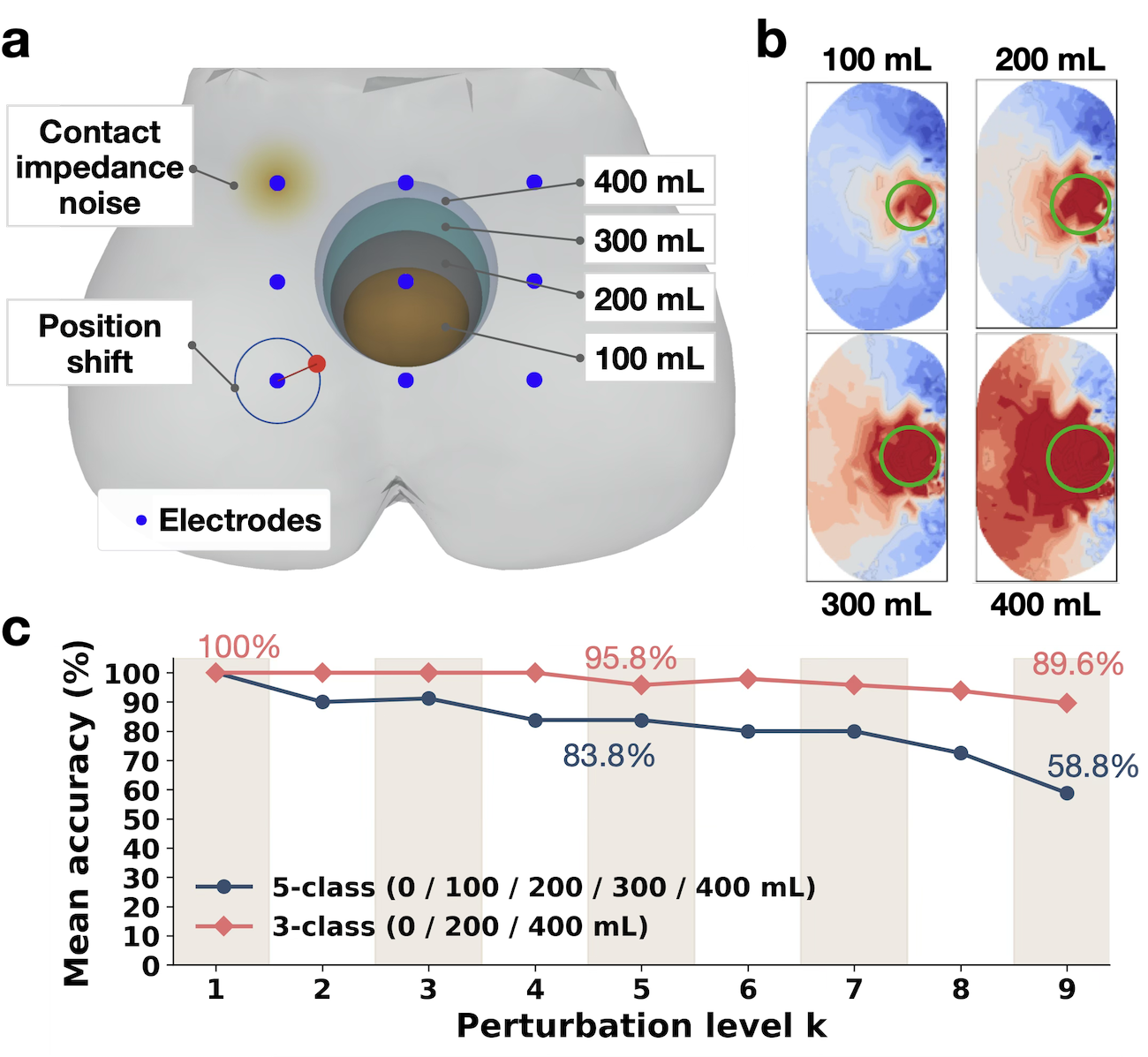}
    \vspace{-2em}
    \caption{Effect of bladder volume under simulated electrode noise, movement \red{and perturbation}. (a) Simulated bladder at different volumes and electrode perturbations. (b) Reconstructed conductivity images at different volume levels. The green circle indicates the area of the bladder. \red{(c) Volume classification accuracy for different volume class under simulated electrode perturbations. }}
    \label{fig:sim_vol1}
    \vspace{-1em}
\end{figure}

\noindent {\bf Image reconstruction.} The raw channel measurements are 1-D time-domain voltage signals, but bladder volume is inherently spatial. We reconstruct a 3D conductivity volume to visualize volume changes. \red{For each  channel, we average over a two-second window to suppress noise, then remove the session baseline by subtracting the first frame from all subsequent ones. From these values we reconstruct the 3D relative-conductivity change using PyEIT's Jacobian-based algorithm (regularization $p=0.5$ for spatial smoothing) and take a 2D slice at the height of the bladder center.}

\subsubsection{Effect of electrode layout.} The electrode layout is an important design choice that dictates \red{the pad's physical size and the number of contacts that must stay reliable against the skin.} We therefore characterized two parameters that affect system performance: \red{the electrodes count and configuration}, and the spacing between them. 

For each electrode design, we simulated a bladder volume of 100~mL, and quantified how well the reconstructed signal was localized to the bladder. 
Specifically, we computed the \textit{RoI response ratio}, the mean absolute reconstructed response inside the bladder region divided by that over the entire domain. Higher values indicate that the reconstructed response is on average stronger within the target bladder region than in the overall background.


\red{We compared four electrode layouts and three spacings as shown in Fig.~\ref{fig:sim_electrodes}. More electrodes consistently produced higher RoI response ratios, from 1.91 ($4\times2$) to 3.83 ($3\times3$), 8.15 ($3\times4$), and 8.46 ($4\times4$), and wider spacing also improved RoI response ratio from 1.81 (30~mm) to 2.88 (45~mm) and 3.83 (60~mm).}


Notably, though EIT is able to sense the conductivity change even with a small grid, a larger area of coverage provides improved spatial resolution across the entire region of interest. For our final hardware design, we select a large $3 \times 3$ grid as a practical compromise to emphasize portability and makes it easier to ensure reliable contact across all electrodes.







\begin{figure*}[t]
\includegraphics[width=\linewidth]{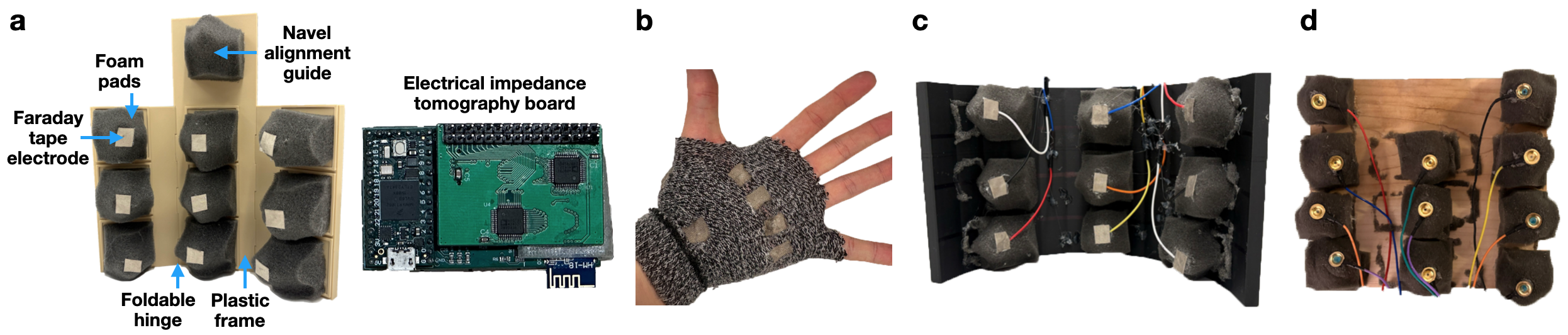}
\vspace{-2em}
\caption{{\bf (a)} Handheld and conformable hardware prototype for bladder fullness sensing using EIT. Alternative prototypes explored: (b) Glove form factor (c) Rigid curved frame (d) Gold-plated electrodes.}
\label{fig:hardware1}
\vspace{-1em}
\end{figure*}

\begin{figure*}[ht]
\includegraphics[width=\linewidth]{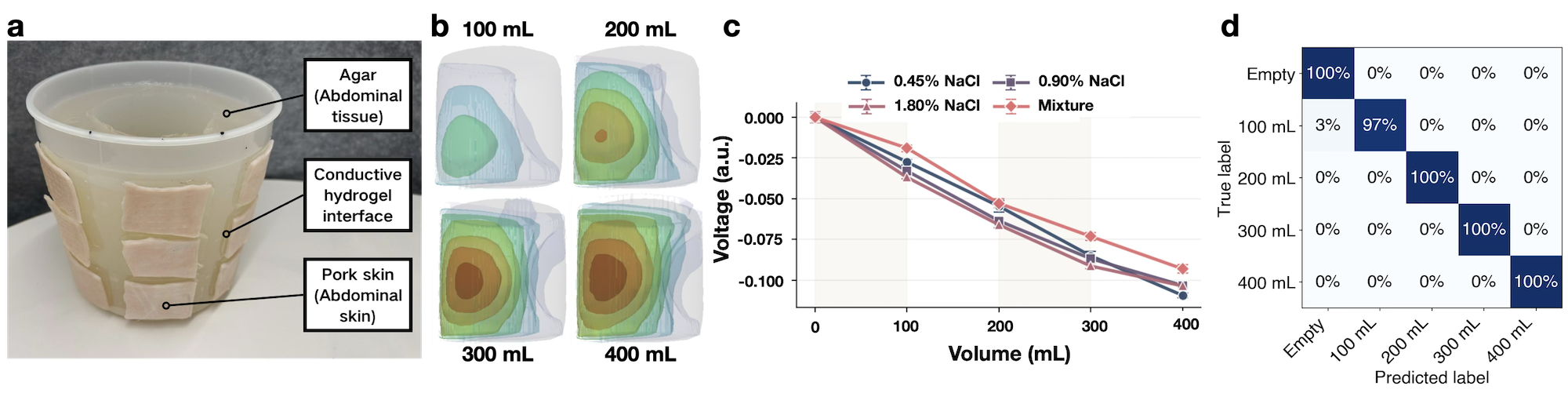}
\vspace{-2em}
\caption{Benchtop validation on bladder phantom. {\bf (a)} Abdomen simulating tissue, bladder, and skin. {\bf (b)} Reconstructed conductivity images from 100 to 400~mL. {\bf (c)} Mean voltage decreases with bladder volume with similar trends across different urine salinities. {\bf (d)} Confusion matrix for volume classification across different urine salinities.
} 
\label{fig:phantom1}
\vspace{-1em}
\end{figure*}

\subsubsection{Effect of electrode perturbations.} In the real-world electrodes may make uneven skin contact and differ slightly in placement between sessions. We therefore evaluate how well the system can distinguish between different urine volume levels under electrode noise and placement error (Fig.~\ref{fig:sim_vol1}).

We treat 400~mL as the upper bound of our sensing range, since it's typically the level at which the urge to void becomes strong~\cite{whec2009urodynamic}. We divide this range into evenly spaced bins, yielding a 3-class problem (0, 200, 400~mL) and a 5-class problem (0, 100, 200, 300, 400~mL).



\red{We apply two types of perturbation. First is impedance shifts scaling the surface impedance near electrodes to 2--5$\times$ nominal. Second is electrode relocations of 5--20~mm modeling placement error. \textit{Both perturbation levels are intentionally large to stress-test the system beyond realistic conditions.} A degree $k$ sets how many of the nine electrodes are randomly perturbed.  we sweep $k$ from 0 to 9 and take 16 independent measurements per volume level.}

\red{We then perform volume classification with a leave-one-level-out scheme on both the 3-class and 5-class problems.} Fig.~\ref{fig:sim_vol1}c show that with no perturbation ($k=0$), we achieve 100\% accuracy for both class divisions. For the 3-class division, this perfect accuracy holds until $k=5$ where it decreases to 95.8\%, and reaches 89.6\% at $k=9$. For the 5-class division, the decline is steeper, falling to 83.8\% at $k=5$ and 58.8\% at $k=9$. 




\subsection{Benchtop validation}

\subsubsection{Hardware design}\leavevmode


\noindent {\bf Sensor board.} Our system (Fig.~\ref{fig:hardware1}a) builds on EIT-kit~\cite{eitkit2021}, which acquires bioimpedance measurements from multiple electrodes in a four-terminal measurement configuration. We use an excitation signal at a frequency of 50~kHz and a sampling rate of 3~Hz which is typically used for bladder sensing~\cite{li2016preliminary,schlebusch2014bladder}.

\noindent {\bf Sensing pad.} We use fabric Faraday tape electrodes~\cite{amazon:faradaytape} mounted on compressible foam pads~\cite{amazon:eggcratefoam}, which provide more flexible placement than rigid electrodes, arranged in a 3$\times$3 grid. To adapt to different abdomen geometries, the foam pads are attached to a foldable 3D-printed plastic frame with hinged joints that allows the system to conform to curved surfaces. For consistent placement, an extra foam pad above the grid is aligned to the navel, which provides a fixed anatomical reference across repeated measurements. 

\subsubsection{Prior design iterations.} During the hardware design process, we also explored other form factors as shown in Fig.~\ref{fig:hardware1}b--d, but chose not to go forward with them for different reasons. \emph{Form factor:} we investigated the use of a glove form factor, but found it was harder to maintain consistent and stable contact across electrodes. \emph{Frame:} we investigated a rigid plastic frame with a fixed curvature, but it did not adapt well to different abdomens shapes which negatively affected electrode contact and signal quality. \emph{Electrodes:} we investigated the use of gold-plated electrodes, but found they did not provide better contact performance than the fabric electrodes. We opted for the fabric electrodes, as they are more conformable and comfortable against the skin, without the cold feel of metal.

\begin{figure*}[t]
    \centering
    \includegraphics[width=\linewidth]{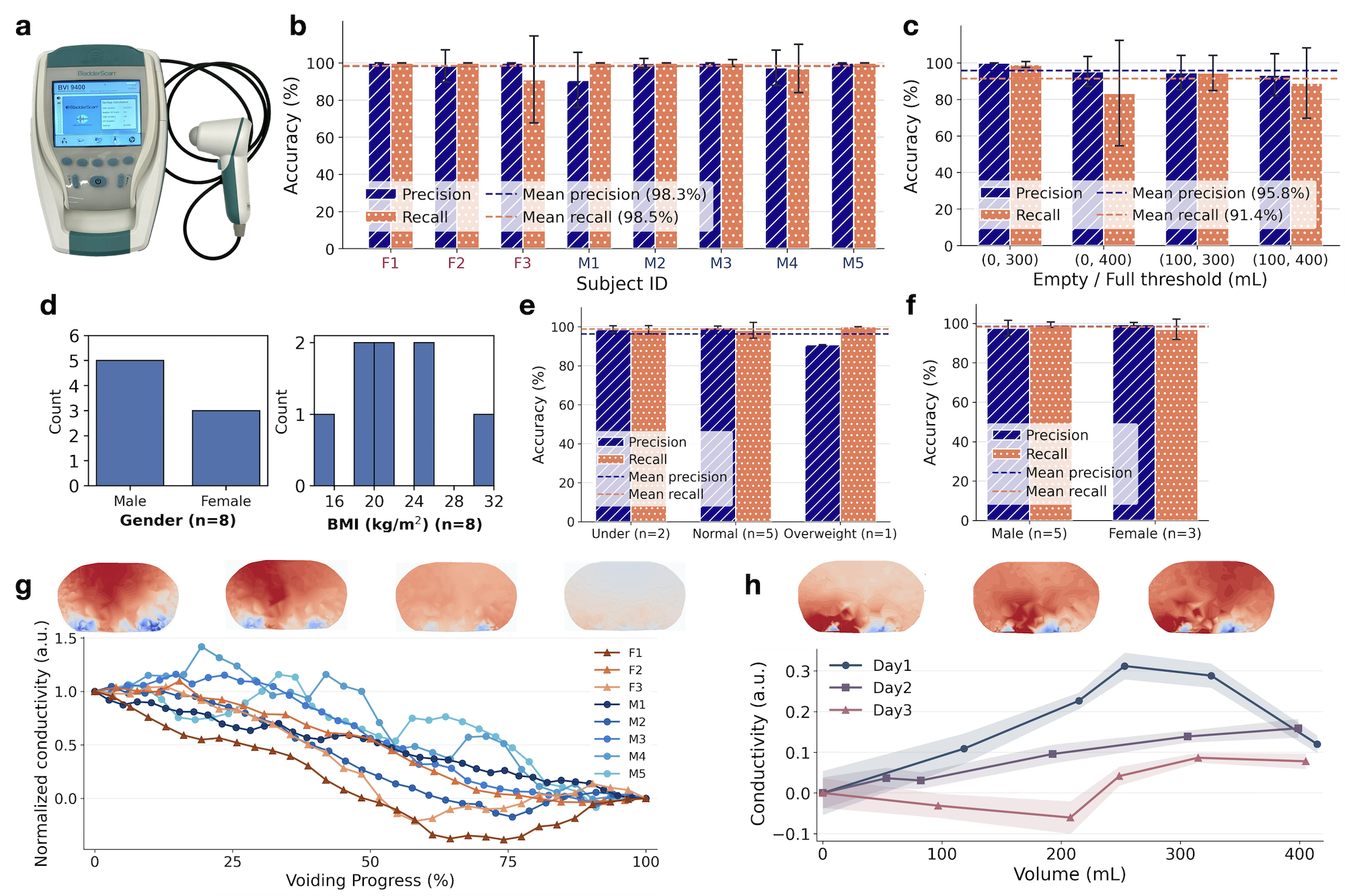}
    \vspace{-2em}
  \caption{In-vivo evaluation. {\bf (a)} Bladder ultrasound~\cite{bvi9400} used for ground truth measurement of bladder volume \red{{\bf (b)} Precision and recall of bladder-fullness detection across 8 subjects (F1--F3, female; M1--M5, male). {\bf (c)} Precision and recall of bladder-fullness detection on 1 subject across days and across empty/full volume-threshold definitions (mL). {\bf (d)}  Demographic summary of participants in the invivo study. {\bf (e)}  Subgroup analysis on bladder fullness detection performance by gender. {\bf (f)}  Subgroup analysis on bladder fullness detection performance by BMI.
  {\bf (g)} Voiding curve across 8 subjects.} {\bf (h)} Filling curve on three different days, each captured over the course of several hours.}
  \label{fig:invivo}
  \vspace{-1em}
\end{figure*}

\subsubsection{Phantom design}

We create a benchtop phantom to evaluate our system (Fig.~\ref{fig:phantom1}a,b). To simulate abdominal tissue, we prepared a bowl of agar by boiling 2~L of tap water in a water distiller~\cite{amazon:cozwaterdistiller} and mixed in 16~g of agar powder~\cite{amazon:agar} and 2~g of NaCl, which has a permittivity of ~$0.2 S\cdot m^{-1} $, similar as soft tissue~\cite{tfdEIT2025}. Once set, we hollowed out a 500~mL cavity at the center to represent the bladder. At each electrode position, we cut an opening in the bowl wall and applied a conductive hydrogel film~\cite{amazon:gelpads} as a coupling interface, then layered pork skin over it to emulate abdominal skin.

\subsubsection{Phantom validation results.} \red{We validated the phantom by comparing its channel-wise voltage responses against our matching cylinder-mesh simulation, and find the cosine similarity across channels is 0.990 and Pearson's correlation 0.971, showing a high level of similarity.}

Then we investigate the effect of urine salinity on system performance given its effect on impedance~\cite{salinity1999}. We use 0.45\%, 0.9\%, and 1.8\% NaCl water to simulate different salinities, in line with known values~\cite{perry2026eit,simavita2017}. \red{We take measurements for uniform and mixed salinities across different volumes from 0 to 400 mL, taking 8 measurements per configuration.}


As shown in Fig.~\ref{fig:phantom1}c,d, the average voltage amplitude across channels decreases with volume for all tested salinity conditions. We then perform volume classification using leave-one-salinity-out, and obtained a classification accuracy of 0.994 $\pm$ 0.013. This shows that our system is robust to realistic variation in urine salinity concentration.

\subsection{In-vivo evaluation }


\subsubsection{Data collection protocol.}

\red{We evaluated our system in two parts. The first part collected continuous \textit{voiding curves} and repeated full-vs-empty measurements from 8 participants (5 male, 3 female), aged 20--27, with BMI ranging from 15.6 to 30.8. Each participant was asked to arrive with a full bladder. After a brief training on using the EIT pad, they took 8 measurements while full, a continuous curve while voiding, and another 8 measurements after voiding, all self-administered. The pad was lifted and replaced between every measurement. }

\red{The second part is a longer bladder-filling protocol, carried out with one participant (male, aged 21, BMI 19.6) and measured against ultrasound reference on three separate sessions, one session per day. To suppress measurement noise, every reading in this protocol is repeated. Each ultrasound reading averages 8 scans with the highest and lowest discarded, and each of our {\sysname} device reading consists of 12 repeated measurements, with the pad lifted and re-aligned to the navel landmark between them. The participant avoided eating for two hours before each session and emptied their bladder on arrival, where we recorded both readings as the session baseline. The participant then refilled their bladder gradually by drinking water, with ultrasound tracking the volume every 10 minutes and {\sysname} readings taken every 20 minutes, or whenever the volume approached $200$~mL or $400$~mL. After the final full-bladder measurement, the participant voided, and we took a final measurement of the empty bladder. }Between measurements, the participant moved freely; all measurements were taken while standing.

\subsubsection{Results.}\leavevmode

\noindent \textbf{Bladder fullness detection.} \red{We first evaluated whether the system distinguishes full from empty across all 8 participants, using the raw 48-channel measurements. We train a separate model per participant, holding out 2 of the 8 full and 2 of the 8 empty measurements as a test set, and average results over all such splits. A $z$-scored ridge classifier reaches a mean precision of 98.3\% $\pm$ 3.2\% and a mean recall of 98.5\% $\pm$ 3.2\%, corresponding to a mean accuracy of 97.9\% $\pm$ 2.6\%, with every participant above 93.1\% (Fig.~\ref{fig:invivo}b).}

\red{We then performed a subgroup analysis on the bladder fullness detection results to evaluate system performance across demographic dimensions of sex and BMI (Fig. ~\ref{fig:invivo}d). 
Among the 8 participants, 25.0\% ($n=2$) of subjects were underweight ($BMI < 18.5$), 62.5\% ($n=5$) had normal weight ($18.5 \leq BMI < 24.9$), 12.5\% ($n=1$) were overweight ($BMI \geq 25.0$). For the underweight group, the system achieved average fullness detection accuracy of $98.2\% \pm 2.5\%$. In the normal weight group, the detection accuracy was $98.8\% \pm 1.9\%$. Among overweight subjects, results were $93.1\% \pm 11.2\%$.  (Fig. \ref{fig:invivo}e)
In our study, 37.5\% ($n=3$) subjects were female, and 62.5\% ($n=5$) subjects were male. Performance is comparable between sexes, at 98.0\% $\pm$ 2.3\% accuracy for female and 97.9\% $\pm$ 3.1\% for male participants ((Fig. \ref{fig:invivo}f). The bladder signal is therefore recoverable in both sexes and across the full BMI range tested. }

\red{The volume at which a bladder counts as full is not a fixed quantity, as the point where the urge to void becomes strong varies between individuals~\cite{whec2009urodynamic}. We therefore tested how sensitive fullness detection is to this definition, using data collected from one participant across three days. We evaluated four different empty/full volume thresholds, training on two days and testing on the held-out day. Averaged over these thresholds, the system reaches a precision of 95.8\% $\pm$ 2.9\% and a recall of 91.4\% $\pm$ 6.8\% (Fig.~\ref{fig:invivo}c), indicating that fullness detection generalizes across days and does not rely on one particular choice of threshold.}

\noindent \textbf{Voiding dynamics.} \red{Beyond empty/full binary classification, we examined the continuous conductivity changes during voiding. We show in Fig.~\ref{fig:invivo}g the change in conductivity during voiding for all 8 participants, averaged across channels and downsampled to 1~Hz. To account for individual variations in absolute signal magnitude and voiding time, each curve is amplitude-normalized (pre-void to 1, post-void to 0) and temporally normalized to 100\% duration.
The reconstructed 2D conductivity maps above the plot for one participant show the bladder region becoming progressively less conductive as voiding proceeds. For all 8 participants, the voiding curves consistently exhibit an overall decreasing trend in conductivity, reflecting bladder emptying, with the decrease concentrated in the middle of the voiding event. Some subjects' voiding curve (F2, M1, M2, M3) decline almost linearly, tracking a straight line with $R>0.93$, while some curves show slight excursions away from this trend. We attribute these to artefacts from involuntary body movement during micturition, where abdominal muscle contraction and relaxation produce impedance changes comparable in magnitude to the bladder's own~\cite{patternSNR2018}. To quantify consistency across participants, we computed the Pearson correlation between each pair of normalized curves, yielding an average $R=0.87$. Although not every curve is linear over the course of voiding, the pre-void and post-void endpoints remain clearly separated in every session. The system is therefore still able to capture the full and empty states in the presence of body motion.}


\noindent \textbf{Filling dynamics.} \red{Finally, we investigated the conductivity changes during gradual bladder filling, using data from the single participant with ultrasound volume references. We show in Fig.~\ref{fig:invivo}h the conductivity changes during the filling protocol across three days, along with reconstructed 2D conductivity maps for Day~1.}
We observe that there is a general upward trend, however it is not completely monotonic. During a voiding event, which occurs in seconds, the impedance change can be attributed predominantly to the bladder, yielding a relatively clean monotonic decrease in conductivity. However, filling can occur over the span of hours during which muscle tension, intestinal motion, gas, and posture all shift the measured impedance~\cite{artefact2023, patternSNR2018, BIAfd2017}. As the bladder's contribution is entangled with these competing effects, the filling curve is not completely monotonic. For these reasons, we scope {\sysname} to bladder fullness sensing rather than absolute volume estimation, which still answers the clinically relevant question of when catheterization is needed.
\section{Limitations and discussion}

\noindent {\bf Evaluation on older subjects.} We present a proof-of-concept design of a handheld pad for bladder fullness sensing evaluated in-silico, in-vitro, and in-vivo. Further evaluation should include a broader population, particularly older adults. Urinary incontinence is common in older adults and is often linked to bladder dysfunction. 


\noindent \red{{\bf Per-user calibration.} Abdominal impedance depends on body composition and pelvic anatomy, which differs from one individual to another, so the same change in bladder volume produces different measured responses across users as documented in prior EIT bladder studies~\cite{nonConsistent2011}. Bladder volume is therefore conventionally estimated from a calibration curve fitted to each subject from an initial measurement cycle~\cite{patternSNR2018}. We follow this established approach by acquiring a one-time ultrasound reference when the user’s bladder is full, recording the corresponding bladder volume, and storing the associated impedance measurement as the user’s calibration profile.}

\noindent \red{{\bf Relation to practical deployment.} Reliable operation of {\sysname} currently depends on navel-guided placement, conductive gel at the electrode interface, and consistent skin contact. These requirements are similar to those of ultrasound bladder scanners in existing clinical practice which also rely on gel coupling and operator-guided positioning. Yet unlike a rigid ultrasound probe, {\sysname} offers a soft pad that conforms to the body and supports on-demand measurement. Quantifying robustness under routine repeated use together with lightweight usability aids such as a real-time contact-quality indicator, is a future step toward deployment in care settings.}





\end{document}